\documentclass[12pt]{iopart}
\usepackage{iopams}
\usepackage{dsfont}
\usepackage[dvips]{graphicx}
\usepackage{setspace}
\newcommand{\half}{\mbox{$\textstyle \frac{1}{2}$}}

\newcommand{\re}{\mbox{$\rm e$}}
\newcommand{\ri}{\mbox{$\rm i$}}
\newcommand{\rd}{\mbox{$\rm d$}}

\begin{document}
\newtheorem{prop}{Proposition}

\title[Coherent states and rational surfaces]{Coherent states and
rational surfaces}

\author[D~C~Brody \& E~M~Graefe]{Dorje~C~Brody and Eva-Maria Graefe}

\address{Department of Mathematics, Imperial College London,
London SW7 2AZ, UK \\ 
School of Mathematics, University of Bristol, Bristol BS8 1TW, UK}


\begin{abstract}
The state spaces of generalised coherent states associated with 
special unitary groups are shown to form rational curves and 
surfaces in the space of pure states. These curves and surfaces 
are generated by the various Veronese embeddings of the 
underlying state space into higher-dimensional state spaces. This 
construction is applied to the parameterisation of generalised 
coherent states, which is useful for practical calculations and provides 
an elementary combinatorial approach to the geometry of the 
coherent state space. The results are extended to Hilbert 
spaces with indefinite inner products, leading to the introduction 
of a new kind of generalised coherent states. 
\end{abstract}


\submitto{\JPA}
%
\vspace{0.4cm}


\noindent 1. \textit{Introduction}.
In the present paper we show that the space of generalised 
coherent states associated with the group $SU(k+1)$ for any 
$k=1,2,\ldots$ can be precisely characterised through the 
algebraic-geometric concept of a Veronese variety, which 
concerns certain embeddings of projective spaces into those 
of higher dimension. This formulation elucidates the geometry 
of $SU(k+1)$ coherent states in an elementary and visual 
manner. This previously unobserved link between the 
geometry of rational curves and surfaces and the theory of 
generalised coherent states constitutes a striking example of 
the intrinsic geometrical aspects of quantum theories and 
may furthermore be useful for practical calculations. 

The paper begins with an introduction to $M$-mode 
Glauber coherent states and generalised $SU(M)$ 
coherent states. A rearrangement of terms demonstrates 
how these two concepts are related. This is followed by 
a brief introduction to the geometry of complex projective 
spaces and the associated Fubini-Study metrics. We then 
discuss the Veronese embeddings of a complex projective 
line into higher-dimensional complex projective spaces, and 
analyse the geometry of algebraic curves obtained from these 
embeddings. Further properties of Glauber coherent states 
are then presented in the style of Geroch (1971), as also used 
in Field \& Hughston (1999) and Brody \& Hughston (2000), 
whereby the connection with the theory of rational curves and 
Veronese embeddings becomes apparent. Glauber 
coherent states are then compared with $SU(2)$ coherent states. 
We show, in particular, how $SU(2)$ coherent states
form rational curves in complex projective spaces.
This is followed by an analysis of $SU(k+1)$ coherent states
for $k=1,2,\ldots$, which correspond to algebraic varieties 
associated with generalised Veronese embeddings. This method 
permits the explicit parameterisation of arbitrary $SU(k+1)$ 
coherent states in arbitrary dimensions as well as the 
geometrical description of coherent state spaces in a 
straightforward manner. We then consider the $SU(1,1)$ 
coherent states of Solomon (1971) and Perelomov (1972), 
and derive the hyperbolic 
metric induced on the coherent state manifold by the 
ambient Fubini-Study geometry. Finite-dimensional analogues 
of $SU(1,k)$ coherent states are then introduced. These 
coherent states appear naturally in the context of Hilbert 
spaces endowed with indefinite inner products. Our discussion 
of coherent states will be primarily focussed upon their 
geometrical aspects; for a general exposition of coherent 
states, see, e.g., Klauder \& Sudarshan (1968); Klauder \& 
Skagerstam (1985); Perelomov (1986); Zhang \textit{et al}. 
(1990); Berman \textit{et al}. (1994); and Vourdas (2006).

\vspace{0.4cm} 
\noindent 2. \textit{Quantum coherent states}. 
In quantum mechanics the state of a single-particle system 
is characterised by a vector in Hilbert space ${\mathcal H}$ 
equipped with a Hermitian inner product, while observables 
are represented by self-adjoint operators acting on 
${\mathcal H}$. The single particle Hilbert space may be 
finite- or infinite-dimensional. The dynamics of the quantum 
system described by a 
Hamiltonian operator $\hat H$ is governed by the Schr\"odinger 
equation, whose solution is given by the action of the time 
evolution operator $\hat U(t)={\exp}(-\rmi\hat H t/\hbar)$ on the 
initial state, where $\hbar$ denotes the Planck constant. 

In many applications of quantum mechanics, subspaces of 
quantum states possessing certain physical properties are of 
particular interest. An important example is that of  
coherent states, which satisfy minimal uncertainty 
conditions and are such that the `classical' dynamics of the 
system is determined by the leading-order dynamics of these 
states in an expansion in powers of $\hbar$. The 
coherent state concept in its modern form was introduced by 
Glauber (1963) in the context of quantum optics for multi-boson 
systems, which we shall discuss briefly below. 

Given a single particle Hilbert space ${\mathcal H}$, the state 
space of a general multi-particle system can be constructed as 
follows.  
The Hilbert space for a combined system of two particles is the 
tensor product ${\mathcal H} \otimes {\mathcal H}$, symmetrised 
for bosons and antisymmetrised for fermions, and similarly for 
three and more particles. The direct sum of these multi-boson 
state spaces forms a Fock space
\begin{eqnarray}
{\mathcal F} = {\mathds C} \oplus {\mathcal H} \oplus ({\mathcal H}
\otimes_s {\mathcal H}) \oplus ({\mathcal H} \otimes_s {\mathcal H}
\otimes_s {\mathcal H}) \oplus \cdots . \label{eq:xx9}
\end{eqnarray}
Here $\otimes_s$ denotes symmetrised tensor product. A 
convenient basis for the Fock space arising from an $M$-dimensional 
single particle Hilbert space ${\mathcal H}$ is the so-called Fock 
basis {$| n_1 \ldots n_M\rangle$}, where $n_j$ denotes the number 
of particles in the $j$th basis state of ${\mathcal H}$. 
A generic state vector $|\xi\rangle$ in the Fock space can then 
be expressed in the form
\begin{eqnarray}
|\xi\rangle = \sum_{n_1\cdots n_M}  \xi_{n_1\ldots n_M}|n_1\ldots 
n_M\rangle.
\end{eqnarray} 
We define $M$ pairs of ladder operators $\{\hat A_j\}$ and 
$\{\hat A_j^\dagger\}$, annihilating or creating a particle in the 
$j$-th basis state: 
\begin{eqnarray}
\left\{ \begin{array}{l} {\hat A^\dagger}_j | n_1 \ldots n_M\rangle = 
\sqrt{n_j+1}\,| n_1\ldots n_j\!+\!1\ldots n_M\rangle, \\ 
{\hat A}_j | n_1 \ldots n_M\rangle = \sqrt{n_j}\,| n_1
\ldots n_j\!-\!1\ldots n_M\rangle. \end{array} \right. 
\label{eq:creat}
\end{eqnarray}   
These operators satisfy the canonical commutation relations 
$[\hat A_j, \hat A_k^\dagger]=\delta_{jk}$ and 
$[\hat A_j, \hat A_k]=[\hat A_j^\dagger, \hat A_k^\dagger]=0$, 
and form the Lie algebra of the Heisenberg-Weyl group. 
(Note that these creation and annihilation operators should 
not be confused with the multiparticle creation and annihilation 
operators used, e.g., in Katriel \textit{et al}. 1987.) 

The $M$-mode Glauber coherent states $|a\rangle$ can be 
defined via the action of a displacement operator 
$\hat D(a)=\prod_j\rme^{a_j\hat A_j^\dagger-a_j^*\hat A_j}$ 
of the Heisenberg-Weyl group on the vacuum state $|0\rangle
=|0\ldots0\rangle$, in which none of the single-particle states is 
populated:
\begin{eqnarray}
|a\rangle&=&\hat D(a)|0\rangle=\prod_j\rme^{a_j\hat A_j^\dagger-
a_j^*\hat A_j}|0\rangle\nonumber\\
&=&\re^{\frac{1}{2}\sum_j |a_j|^2}\!\!\!\!\sum_{n_1,\cdots, 
n_M=0}^\infty\!\left(\,\prod_{j=1}^M \frac{a_j^{n_j}}{\sqrt{n_j!}}
\right)|n_1\ldots n_M\rangle.
\label{eq:GCS}
\end{eqnarray}
Such states possess a number of special features and therefore play 
important roles in various fields of quantum physics. 
First, they are eigenstates of the annihilation operators, i.e. 
${\hat A}_j|a\rangle=a_j|a\rangle$. Second, they form an 
`over-complete' set of basis vectors for the symmetric Fock
space and constitute a resolution of the identity. (The latter property 
is related to the 
fact that projective varieties associated with coherent state 
manifolds are `balanced' in the ambient state space manifold; 
see Donaldson 2001). Third, they saturate the lower 
bound of the position-momentum uncertainty relation (where the
position and momentum operators are the real and imaginary 
parts of the annihilation operator, respectively), and thus 
represent, in some respects, states that are closest to being 
classical. Finally, under time evolution with a Hamiltonian 
that is linear in the generators of the 
Heisenberg-Weyl algebra, an initially coherent state remains 
coherent, and the expectation values of the position and 
momentum operators evolve according to the corresponding 
classical equations of motion. This last property remains valid 
to leading order in $\hbar$ for an arbitrary Hamiltonian, which 
constitutes another reason why coherent states are often 
viewed as representing classical states. 

The algebraic characterisation of Glauber 
coherent states presented above has been generalised to 
systems with arbitrary Lie group structures by Perelomov 
(1972); see also Radcliffe (1971) and Gilmore (1972). 
One important example is that of the $SU(2)$ or so-called 
atomic coherent states. These arise naturally in the context of  
rotationally invariant systems, where the components 
$\hat L_{x,y,z}$ of the angular momentum operator generate 
an algebra isomorphic with $\mathfrak{su}(2)$. If the Hamiltonian 
commutes with the total angular momentum operator 
${\hat{{\mathbf L}}}$, then the Hilbert space of the system is a 
direct sum of joint eigenspaces of ${\hat H}$ and ${\hat L}^2$, 
each rotationally irreducible and of dimension $2L+1$, where 
$L(L+1)$ is the appropriate eigenvalue of ${\hat L}^2$. We 
shall, in the sequel, confine our considerations to one such 
eigenspace. In analogy with 
the Glauber coherent states (\ref{eq:GCS}) the $SU(2)$ coherent 
states $|\theta,\phi\rangle$ can be defined by the action of the 
$SU(2)$ displacement operator $\hat D(\theta, \phi)=
\rme^{\rmi\theta(\hat L_x \sin\phi-\hat L_y\cos\phi)}$ on the 
eigenstate $|-L\rangle$ of the angular momentum operator 
$\hat L_z$ corresponding to the lowest eigenvalue:
\begin{eqnarray}
|\theta,\phi\rangle=\hat D(\theta,\phi)|-L\rangle=
\rme^{\rmi\theta(\hat L_x\sin\phi-\hat L_y\cos\phi)}|-L\rangle.
\label{eq:SU2CS}
\end{eqnarray}
These atomic coherent states satisfy the minimal 
uncertainty relations for the angular momentum operators, 
and their dynamics coincide with those of the corresponding 
classical states to leading order in $\hbar$. 

Interestingly, the Glauber coherent states for 
an $M$-dimensional single particle Hilbert space can be 
constructed as a sum over $SU(M)$ coherent states, as we 
shall briefly demonstrate below for the two-dimensional case. 
In accordance with the Schwinger representation of angular 
momentum (Schwinger 1952), the angular momentum operators 
may be described in terms of a two-state Heisenberg-Weyl 
algebra spanned by $\hat A_{1}$, $\hat A_{2}$ and 
$\hat A_{1}^\dagger$, $\hat A_{2}^\dagger$ as 
\begin{eqnarray}
\fl L_x=\half(\hat A_1^\dagger\hat A_2+\hat A_2^\dagger\hat A_1), 
\quad 
L_y={\mbox{$\textstyle \frac{1}{2\rmi}$}}(\hat A_1^\dagger\hat A_2-
\hat A_2^\dagger\hat A_1), \quad 
L_z=\half(\hat A_1^\dagger\hat A_1-\hat A_2^\dagger\hat A_2)
\end{eqnarray}
with the additional restriction that the number of bosons is fixed 
as $N=2L$. Thus, we can interpret a $(2L+1)$-dimensional 
$SU(2)$ system as a two-state system populated with $2L$ 
bosons, and in the Fock basis the $SU(2)$ coherent states 
(\ref{eq:SU2CS}) may be expressed in the form
\begin{eqnarray}
|\theta,\phi\rangle&=&\hat D(\theta,\phi)|0,N\rangle=
\rme^{\rmi\theta(\hat L_x\sin\phi-\hat L_y\cos\phi)}|0,N\rangle\nonumber\\
&=&\sum_{j=0}^N \sqrt{{\textstyle \left( {N\atop
j} \right)}} \left( \cos\half\theta \right)^{N-j} \left( \sin\half\theta \re^{{\rm i}\phi} 
\right)^{j} |j,N-j\rangle. \label{eq:ACS} 
\end{eqnarray}
On the other hand, the Glauber coherent states for two modes can 
be rewritten as
\begin{eqnarray}
|a_1,a_2\rangle&=&\rme^{\frac{1}{2}(|a_1|^2+|a_2|^2)} 
\sum_{n_1, n_2=0}^\infty \frac{a_1^{n_1}}{\sqrt{n_1!}}
\frac{a_2^{n_2}}{\sqrt{n_2!}}|n_1, n_2\rangle\nonumber\\
&=&\rme^{\frac{1}{2}(|a_1|^2+|a_2|^2)} \sum_{N=0}^\infty 
\sum_{j=0}^N \frac{a_1^{N-j}a_2^{j}}{\sqrt{(N-j)!j!}}|j, 
N-j\rangle\nonumber\\
&=&\rme^{\frac{1}{2}(|a_1|^2+|a_2|^2)} \sum_{N=0}^\infty 
\frac{1}{\sqrt{N!}}\sum_{j=0}^N  \sqrt{{\textstyle \left( 
{N\atop j} \right)}}a_1^{N-j}a_2^{j}|j, N-j\rangle.
\end{eqnarray}
Hence, identifying $a_1=c\cos\half\theta$ and 
$a_2=c\sin\half\theta \re^{{\rm i}\phi}$ with $c\in{\mathds C}-\{0\}$, 
we see that the terms in the Glauber coherent states with constant 
boson number $N$ are proportional to the $SU(2)$ coherent states. 
The relation between $M$-mode Glauber coherent states and 
$SU(M)$ coherent states for arbitrary $M$ can be established by 
an analogous construction. 

In view of the important role played by coherent states in various 
physical applications, we shall analyse in further detail the geometry 
of the subspaces of the quantum state space spanned by the 
coherent states. Before proceeding, however, we first briefly review 
the Fubini-Study geometry of the space of pure states, and the 
concepts of rational curves and Veronese embeddings.

\vspace{0.4cm} 
\noindent 3. \textit{Fubini-Study geometry of quantum state space}. 
In quantum mechanics the expectation value of an observable 
${\hat H}$ in a state $|z\rangle\in{\mathcal H}$, which represents 
the outcome of measurements, is given by 
$\langle z,{\hat H} z\rangle/\langle z,z\rangle$. This is invariant 
under the overall complex phase shift $|z\rangle \to \lambda|z
\rangle$, $\lambda\in{\mathds C}-\{0\}$. Hence two 
Hilbert space vectors only differing by a complex scale factor 
represent the same physical state. We can therefore eliminate 
the overall physically irrelevant degree of freedom by considering 
the space of equivalence class modulo the identification $|z\rangle 
\sim \lambda|z\rangle$. The resulting projective Hilbert space is 
the space of quantum states, which, in the case of a finite, 
($n+1$)-dimensional system is the complex projective $n$-space 
$\mathds{CP}^n$. The geometry of the space of pure states, the 
Fubini-Study geometry, characterises important aspects of the 
physical behaviour of a system, and will be reviewed in the following. 

We begin by remarking that the complex projective space 
$\mathds{CP}^n$ is the quotient space of a real $2n+1$ sphere 
by the circle group $U(1)$, i.e. $\mathds{CP}^n=
S^{2n+1}/U(1)$. This can be seen as follows. 
Consider a complex projective line $\mathds{CP}^1$, 
i.e. the quotient space of ${\mathds C}^2$ under the
identification $(w_1,w_2)\sim(\lambda w_1,\lambda w_2)$ for all
$\lambda \in {\mathds C}-\{0\}$. In other words, $\mathds{CP}^1$ is
the space of rays through the origin of ${\mathds C}^2$; two points
in ${\mathds C}^2$ define the same point of $\mathds{CP}^1$
iff they lie on the same complex line through the origin. Now
if we first normalise the overall scale of ${\mathds C}^2$, we
obtain the three sphere $|w_1|^2+|w_2|^2=1$; if we further 
identify vectors differing only by a phase factor,
then we obtain the Hopf fibration $S^3/U(1) = S^2$. Thus, in real 
terms the complex projective line can be viewed as a two sphere 
$S^2$. This is also evident from the fact that the
intersection of a sphere and a line through the origin of ${\mathds
C}^2$ is a real circle. Analogous constructions clearly exist in 
higher dimensions. Thus, for ${\mathds C}^3$ we normalise the 
scale to obtain a five sphere $|w_1|^2+|w_2|^2+|w_3|^2=1$, 
whence the complex projective plane
$\mathds{CP}^2$, i.e. the space of rays through the origin of
${\mathds C}^3$, is obtained via the Hopf fibration $S^5/U(1) =
\mathds{CP}^2$. More generally, the space of rays in ${\mathds
C}^{n+1}$ is just the quotient $\mathds{CP}^n=S^{2n+1}/U(1)$,
since every line through the origin of ${\mathds C}^{n+1}$
intersects the unit sphere in a circle.
 
The points of the complex projective space $\mathds{CP}^n$ are 
often conveniently represented by homogeneous coordinates
$(z^0,z^1,z^2,\ldots,z^n)$, with a
redundant complex degree of freedom. To specify the geometry of
$\mathds{CP}^n$ induced by the ambient Euclidean geometry of
${\mathds C}^{n+1}$, consider the inner
product of neighbouring points. Writing $\rd s$ for
the line element and $\langle~,~\rangle$ for the inner product in
${\mathds C}^{n+1}$ we have
\begin{eqnarray}
\frac{\langle {\bar z},z+\rd z\rangle  \langle {\bar z}+ \rd {\bar
z},z\rangle} {\langle {\bar z},z\rangle\langle {\bar z}+\rd{\bar
z},z+\rd z\rangle} = \cos^2\half \rd s. \label{eq:1}
\end{eqnarray}
Solving this for $\rd s$ and retaining terms of quadratic order, 
we obtain the Fubini-Study line element
\begin{eqnarray}
\rd s^2 = 4 \frac{\langle {\bar z},z\rangle\langle \rd{\bar z}, \rd
z\rangle - \langle {\bar z},\rd z\rangle \langle z,\rd{\bar z}
\rangle}{\langle {\bar z},z\rangle^2}. \label{eq:2}
\end{eqnarray}
 
An alternative derivation of the Fubini-Study metric employs the 
fact that complex projective spaces possess K\"ahlerian
structures. In particular, the K\"ahler potential $K$ for
$\mathds{CP}^n$, in terms of inhomogeneous coordinates
$(\zeta^1,\zeta^2,\ldots,
\zeta^n)=(z^1/z^{0},z^2/z^{0},\ldots,z^n/z^{0})$ on an appropriate
coordinate patch $z^0\neq0$, is given by
\begin{eqnarray}
K=4\ln(1+{\bar\zeta}_j\zeta^j). \label{eq:xx3}
\end{eqnarray}
Thus, by use of the standard definition 
\begin{eqnarray}
\rd s^2 = \frac{\partial^2
K}{\partial\zeta^i\partial{\bar\zeta}_j}\, \rd \zeta^i \rd
{\bar\zeta}_j 
\end{eqnarray}
of a K\"ahler metric we obtain the familiar expression (Kobayashi \& 
Nomizu 1969) for the Fubini-Study metric:
\begin{eqnarray}
\rd s^2 =
4\frac{(1+{\bar\zeta}_j\zeta^j)(\rd{\bar\zeta}_j\rd\zeta^j)-
({\bar\zeta}_j\rd\zeta^j)(\zeta^j\rd{\bar\zeta}_j)}
{(1+{\bar\zeta}_j\zeta^j)^2}. \label{eq:xx5}
\end{eqnarray} 
The quantum mechanical transition probabilities are thus measured 
in terms of the Fubini-Study geodesic distances between the states. 
The expression for the metric will be used in what follows to determine 
the induced metrics of the various subspaces of the Fubini-Study 
manifold. In particular, we now turn to the discussion of the Veronese 
embedding in the Fubini-Study manifolds and derive the geometry 
induced by the embedding. 

\vspace{0.4cm} 
\noindent 4. \textit{Veronese embeddings and rational curves}. We 
have seen that the $M$-mode Glauber coherent states can be viewed 
as consisting of a combination of $SU(M)$ coherent states over 
different particle numbers. As we shall indicate later, the state space 
of $SU(M)$ coherent states arises from the Veronese embedding, the 
concept of 
which we shall briefly introduce here. The Veronese variety is 
concerned with the embedding of a complex projective space into 
higher dimensional complex projective spaces, in particular, it is an 
embedding of $\mathds{CP}^k$ in $\mathds{CP}^{k(k+3)/2}$
possessing certain properties (Nomizu 1976). Here we review the 
properties of this embedding for $k=1$, that is, $\mathds{CP}^1 
\hookrightarrow \mathds{CP}^2$ and its generalisations 
$\mathds{CP}^1 \hookrightarrow \mathds{CP}^n$, which 
will be shown to characterise $SU(2)$ coherent states. 

Let $(s,t)$ be the homogeneous coordinates of a point in 
$\mathds{CP}^1$. The image point in $\mathds{CP}^2$ under 
the Veronese embedding then has the homogeneous coordinates
$(s^2,\sqrt{2}st,t^2)$. Thus, the image of $\mathds{CP}^1$ in the 
complex projective plane $\mathds{CP}^2$ forms a conic curve
${\mathcal C}$ characterising the solution to a quadratic equation. 
Because $\mathds{CP}^1$ in real terms is a two-sphere, this 
nonsingular one-to-one correspondence between points on 
$\mathds{CP}^1$ and points on ${\mathcal C}$ implies that the 
conic ${\mathcal C}$ is a topological  sphere. The metric on 
${\mathcal C}$ induced by the ambient Fubini-Study metric of 
$\mathds{CP}^2$ can be calculated as follows. We substitute 
the homogeneous coordinates $(z^1,z^2,z^3) = (s^2,
\sqrt{2}st,t^2)$ of $\mathds{CP}^2$ into formula (\ref{eq:2}) for the
line element, and perform the same calculation for the metric of
$\mathds{CP}^1$ in terms of the homogeneous coordinates
$(z^1,z^2)=(s,t)$. A short calculation then yields
\begin{eqnarray}
\rd s_{\mathcal C}^2=2\rd s_{\mathds{CP}^1}^2,
\end{eqnarray}
that is, the metric of the conic ${\mathcal C}$ is just twice the metric
of $\mathds{CP}^1$. If we fix the scale
so that the radius of $\mathds{CP}^1=S^2$ is one, as we have done
implicitly in (\ref{eq:1}), then in real terms the conic ${\mathcal
C}$ is a two-sphere of radius $\sqrt{2}$.
 
The Veronese embedding of $\mathds{CP}^1$ in $\mathds{CP}^2$ 
can be generalised in a natural way to an embedding of the form
$\mathds{CP}^1 \hookrightarrow \mathds{CP}^n$ such that if $(s,t)$
denotes the homogeneous coordinates for $\mathds{CP}^1$, then 
the correspondence
\begin{eqnarray}
(s,t) \hookrightarrow \left( s^n, \sqrt{{\textstyle \left( {n\atop
1} \right)}} s^{n-1}t, \sqrt{{\textstyle \left( {n\atop 2} \right)
}} s^{n-2}t^2, \cdots, \sqrt{{\textstyle \left( {n\atop n-1} \right)
}} s t^{n-1}, t^n \right) \label{eq:3}
\end{eqnarray}
defines the homogeneous coordinates of the image in 
$\mathds{CP}^n$. The image for $n=3$ is a twisted cubic 
curve (cf. Wood 1913), for $n=4$
a rational quartic curve (cf. Telling 1936), for $n=5$ a
rational quintic curve, for $n=6$ a rational sextic curve, and so on. 
Thus, the generalised Veronese embedding of $\mathds{CP}^1$ in
$\mathds{CP}^n$ defines a rational curve in $\mathds{CP}^n$ (an
algebraic curve in $\mathds{CP}^n$ has topological dimension 2
and thus represents a surface, and rational curves are characterised 
by the fact that the corresponding surfaces have zero genus); the 
significance of these rational curves with respect to the structures 
of elementary quantum spin systems is discussed in detail 
in Brody \& Hughston (2001).
 
Using the representation (\ref{eq:3}), it is not difficult to
specify the geometry of these rational curves. Since direct
computation of the metric is cumbersome, we proceed by first 
calculating the K\"ahler potential (cf. Stanley 1993). We note 
that the K\"ahler potential for $\mathds{CP}^1$ is 
$K_{\mathds{CP}^1}=4\ln(1+|t/ s|^2)$. This follows from
(\ref{eq:xx3}) by setting $\zeta=t/s$ for the inhomogeneous
coordinate of $\mathds{CP}^1$. We now use (\ref{eq:3}) and 
follow the same procedure to derive the K\"ahler potential for 
the rational curve ${\mathcal R}$ in $\mathds{CP}^n$. A short
calculation then yields
\begin{eqnarray}
K_{\mathcal R}=4\ln(1+|t/s|^2)^n , \label{eq:xx8}
\end{eqnarray}
which shows that the metric for the rational curve ${\mathcal R} \in
\mathds{CP}^n$ is $n$ times the metric of a two-sphere with unit
radius. It follows that the rational curve ${\mathcal R} \in
\mathds{CP}^n$ in real terms is a two-sphere with radius $\sqrt{n}$
and constant curvature $2/n$.

\vspace{0.4cm} 
\noindent 5. \textit{The geometry of Glauber coherent states}. 
To analyse the geometry of the Glauber coherent states and their 
relation to $SU(M)$ coherent states it is convenient to characterise 
Fock space operations in the style of Geroch (1971). For this 
purpose we write $\xi\in{\mathds C}$, and $\xi^a$ for an element 
of ${\mathcal H}$ where the abstract index $a$ labels the 
components of the single-particle state in an arbitrary chosen 
orthonormal basis set. An element of ${\mathcal H} \otimes_s 
{\mathcal H}$ can then be written $\xi^{ab}=\xi^{(ab)}$, and so on, 
where round brackets denote symmetrisation over tensor indices. 
A generic state vector $|\xi\rangle$ in Fock space can be written 
in the form
\begin{eqnarray}
|\xi\rangle = (\xi, \xi^a, \xi^{ab}, \xi^{abc}, \cdots ).
\end{eqnarray}
The inner product of a pair of states $|\xi\rangle$ and
$|\eta\rangle$ in Fock space is then 
\begin{eqnarray}
\langle\eta|\xi\rangle = {\bar\eta}\xi + {\bar\eta}_a\xi^a +
{\bar\eta}_{ab}\xi^{ab} + \cdots
\end{eqnarray}
Hence, the norm of a state is $\|\xi\|^2=\langle\xi|\xi\rangle$,  
assumed finite. 

A state $|\xi\rangle$ in Fock space can be augmented by a 
particle in the state $\sigma^a\in{\mathcal H}$ via the action 
of the creation operator ${\hat A}^\dagger_\sigma$, given by 
\begin{eqnarray}
{\hat A}^\dagger_\sigma |\xi\rangle = \left(0, \xi\sigma^a, \sqrt{2}
\xi^{(a}\sigma^{b)}, \sqrt{3} \xi^{(ab}\sigma^{c)}, \cdots \right).
\label{eq:7}
\end{eqnarray}
Note that the general creation operator ${\hat A}^\dagger_\sigma$ 
can be viewed as a linear superposition of the elementary creation 
operators $\{\hat A_j^\dagger\}$ introduced in (\ref{eq:creat}). 
Similarly, the annihilation operator ${\hat A}_\tau$ acts as follows:
\begin{eqnarray}
{\hat A}_\tau |\xi\rangle = \left({\bar\tau}_c\xi^c,
\sqrt{2}{\bar\tau}_c\xi^{ac}, \sqrt{3}{\bar\tau}_c \xi^{abc}, \cdots
\right). \label{eq:8}
\end{eqnarray} 
Formulae (\ref{eq:7}) and (\ref{eq:8}) imply the canonical
commutation relations $[{\hat A}^\dagger_\sigma,
{\hat A}^\dagger_{\sigma'}]= [{\hat
A}_\tau,{\hat A}_{\tau'}]=0$ and $[{\hat A}^\dagger_\sigma,{\hat
A}_{\tau}]=({\bar\tau}_a\sigma^a){\mathds 1}$.

A general multi-particle quantum 
state $|\xi\rangle$ in Fock space is fully characterised by 
an analytic function on the single particle Hilbert space 
${\mathcal H}$ (Bargmann 1961). Specifically, the state 
$|\xi\rangle$ can be fully recovered from the function 
\begin{eqnarray}
\Psi(\eta) = {\bar\xi} + {\bar\xi}_a \eta^a +
{\textstyle\frac{1}{\sqrt{2!}}} {\bar\xi}_{ab} \eta^a \eta^b +
{\textstyle\frac{1}{\sqrt{3!}}} {\bar\xi}_{abc} \eta^a \eta^b
\eta^c + \cdots \label{eq:14x}
\end{eqnarray}
for $\eta^a \in {\mathcal H}$. If the function (\ref{eq:14x}) is of 
the exponential form $\Psi(\eta) = \exp( {\bar\xi}_a \eta^a)$, 
the corresponding state is a Glauber coherent state (\ref{eq:GCS}). 
In this case all the multi-particle components depend upon just one  
single-particle state and in the abstract index notation a coherent 
state has the form
\begin{eqnarray}
|\xi\rangle = \left(1, \xi^a, {\textstyle\frac{1}{\sqrt{2!}}}
\xi^a\xi^b, {\textstyle\frac{1}{\sqrt{3!}}} \xi^a\xi^b\xi^c, \cdots
\right). \label{eq:10}
\end{eqnarray}
Because the norm $\|\xi\|$ of a state $|\xi\rangle$ is not physically 
observable, we may projectivise the Fock space ${\mathcal F}$. 
The projectivised coherent states then form a submanifold of the 
projective Fock space. The geometry of the space of Glauber 
coherent states has been investigated in the literature, and the 
results can be summarised as follows:

\noindent \textbf{Theorem (Provost \& Vallee 1980; Field \& Hughston
1999)}. \textit{The geometry of the coherent state
submanifold of the projective Fock space induced by the ambient
Fubini-Study metric is flat, i.e. complex Euclidean}.
 
\textit{Proof}. The Fubini-Study metric on the projective Fock space
is
\begin{eqnarray}
\rd s^2 = 4 \left[ \frac{\langle \rd\xi|\rd\xi\rangle}{\langle
\xi|\xi\rangle} - \frac{\langle \xi|\rd \xi\rangle \langle \rd\xi|
\xi\rangle} {\langle \xi|\xi \rangle^2} \right]. \label{eq:x2}
\end{eqnarray}
For a coherent state (\ref{eq:10}) a straightforward calculation shows
that $\langle\xi|\xi\rangle = \exp( {\bar\xi}_a\xi^a)$,
$\langle\xi|\rd\xi\rangle = \exp( {\bar\xi}_a\xi^a) {\bar\xi}_a\rd
\xi^a$, and $\langle\rd\xi|\rd\xi\rangle = \exp( {\bar\xi}_a\xi^a)
[\rd{\bar\xi}_a\rd \xi^a + ({\bar\xi}_a\rd \xi^a)(\xi^a\rd{\bar\xi}_a)$]. 
Substituting these expressions into (\ref{eq:x2})
we find that the line element is given by $\rd
s^2=4\rd{\bar\xi}_a\rd\xi^a$. \hspace*{\fill} $\square$
 
The above proof is equivalent to the first of the three proofs given
in Field \& Hughston (1999). We also remark that Provost \& Vallee
(1980) established this result for the case of one-mode Glauber 
coherent states. 
 
From the algebraic definition of coherent states the flatness of the 
Weyl-Heisenberg group manifold seems natural, but from the 
geometric point of view this result is at first surprising, since a 
linear superposition of a pair of coherent states is
incoherent, i.e. the complex projective line passing through a pair
of coherent states in the projective Fock space lies outside the
coherent state submanifold (except for the two intersection points).
The `linearity' of the submanifold of Glauber coherent states, i.e. the 
flatness of the induced metric, must be understood with the 
caveat that although $\xi^a$ and $\lambda \xi^a$ for $\lambda\in
{\mathds C}-\{0\}$ represent the \textit{same} single-particle state 
vector, the coherent states arising from $\xi^a$ and  
$\lambda \xi^a$ represent \textit{different} multi-particle state vectors
(the inner product $\langle\xi|\lambda \xi \rangle=
\langle\xi| \xi \rangle^\lambda$ of the associated coherent states 
agrees with $\langle\xi| \xi \rangle$ iff $\lambda=1$). Thus, the 
Glauber coherent-state
submanifold of the projective Fock space is endowed with the
Euclidean geometry of the underlying single-particle Hilbert space,
not the Fubini-Study geometry of the single-particle state space.
However, the situation is somewhat different for the 
atomic coherent states.

\vspace{0.4cm} 
\noindent 6. \textit{Rational curves and atomic coherent states}. 
As discussed above, the $SU(M)$ coherent states can be viewed 
as the $N$-particle component of an $M$ mode Glauber coherent 
state (\ref{eq:10}), suitably renormalised:
\begin{eqnarray}
|\xi\rangle = \frac{\xi^a\xi^b\cdots\xi^c}{({\bar\xi}_a\xi^a)^{N/2}}.
\label{eq:17}
\end{eqnarray} 
By virtue of the normalisation in (\ref{eq:17}), $SU(M)$ coherent 
states, unlike Glauber coherent states, are such that a pair of 
Hilbert space vectors $\xi^a$ and $\lambda \xi^a$ representing 
the same state vector also represent the same $SU(M)$ coherent 
state. Thus, $SU(M)$ coherent states do not inherit the Euclidean 
geometry of the underlying Hilbert space. 

In the case of the \textit{atomic} or $SU(2)$ coherent states, the 
underlying single particle Hilbert space is two-dimensional. 
Introducing spherical variables for the homogeneous coordinates, 
we write $(s,t)=(\cos\half\theta, \sin\half\theta \re^{{\rm i}\phi})$. 
The atomic coherent state (\ref{eq:17}) 
then lies in the $(N+1)$-dimensional subspace of symmetric state 
vectors in the $2^N$-dimensional Hilbert space. 
Writing $\{|j, N-j\rangle\}_{j=0,\ldots,N}$ for the basis elements 
of this Hilbert space, we recover the familiar expression 
(\ref{eq:ACS}) for the $SU(2)$ coherent state. Clearly, the 
coefficients in the $SU(2)$ coherent state (\ref{eq:ACS}) are 
in bijective correspondence with the components of the rational 
curve (\ref{eq:3}) in $\mathds{CP}^N$. We thus conclude that for each 
$N$ the $SU(2)$ coherent states form a rational curve ${\mathcal R}$ 
in the associated projective state space $\mathds{CP}^N$. 

The geometry of the space of $SU(2)$ coherent states 
is therefore equivalent to that of the rational curve discussed above. In 
terms of the spherical parameterisation the inhomogeneous coordinate 
of $\mathds{CP}^1$ is $\zeta=\tan\half\theta\re^{{\rm i}\phi}$, from which 
the Fubini-Study metric can be calculated via (\ref{eq:xx5}). A short 
calculation then gives the standard Riemannian metric for the unit 
sphere: $\rd s^2 = \rd\theta^2 + \sin^2\theta \rd\phi^2$. Moreover, from 
(\ref{eq:xx8}), the metric of the $(N+1)$-component $SU(2)$ 
coherent state space is given by 
\begin{eqnarray}
\rd s_{\mathfrak{su}(2)}^2 = N(\rd\theta^2 + \sin^2\theta \rd\phi^2).
\end{eqnarray}
This agrees with the result obtained in Provost \& Vallee (1980). 
The coherent state manifold may be regarded as a classical phase 
space, with a constant curvature of $2/N$. 

As mentioned above, there exists an elegant geometric characterisation 
of the various spin states in terms of the properties of rational curves 
and their osculating hyperspaces, the 
spin states being determined by the various intersection points of 
these surfaces (Brody \& Hughston 2001).

\vspace{0.4cm} 
\noindent 7. \textit{Generalised Veronese embeddings
as $SU(k+1)$ coherent states}. We now consider the extension of 
the $SU(2)$ atomic coherent states characterised by rational curves 
to the general $SU(k+1)$ coherent states. Recall that a Veronese 
variety is defined by an embedding of $\mathds{CP}^k$ in 
$\mathds{CP}^{k(k+3)/2}$. For each $k$ this embedding can be 
generalised in a natural manner such that various state spaces of 
$SU(k+1)$ coherent states can be generated systematically. We 
begin with the analysis of $SU(3)$ coherent states. 

For $k=2$ the embedding
$\mathds{CP}^2 \hookrightarrow \mathds{CP}^5$ can be
constructed as follows. Let $(s,t,u)$ be the homogeneous coordinates
of a point in $\mathds{CP}^2$. Then, the homogeneous coordinates of 
the image point in $\mathds{CP}^5$ are $(s^2, \sqrt{2}st,
t^2, \sqrt{2}tu, u^2, \sqrt{2}us)$. This defines a rational
quadratic surface (a real four-dimensional simply connected smooth
manifold) in $\mathds{CP}^5$ with the topology of a
$\mathds{CP}^2$.

As in the case of $\mathds{CP}^1$, the Veronese embedding of
$\mathds{CP}^2$ can also be generalised to embeddings of the form
$\mathds{CP}^2 \hookrightarrow \mathds{CP}^{N(N+3)/2}$,
$N=2,3,\ldots$, such that the homogeneous coordinates of the image of
$(s,t,u)$ are defined by the trinomial expansion. For example,
when $N=3$, the homogeneous coordinates of the image point 
in $\mathds{CP}^9$ are $(s^3, \sqrt{3}s^2t, \sqrt{3}st^2, t^3,
\sqrt{3}t^2u, \sqrt{3}tu^2, u^3, \sqrt{3}u^2s, \sqrt{3}us^2,
\sqrt{6}stu)$.

We now determine the induced metric on the rational surfaces by
calculating the K\"ahler potential. Writing
$(\zeta^1,\zeta^2)=(t/s,u/s)$ for the inhomogeneous coordinates of
$\mathds{CP}^2$, the K\"ahler potential of the
generalised Veronese variety $\mathds{CP}^2 \hookrightarrow
\mathds{CP}^{N(N+3)/2}$ is
\begin{eqnarray}
K = 4 \ln (1+{\bar\zeta}_1\zeta^1+{\bar\zeta}_2\zeta^2)^N .
\end{eqnarray}
Thus, the induced metric on these Veronese varieties is
that of a complex projective plane, multiplied by the scale factor
$N$.

The foregoing discussion shows that for each
value of $N$ the embedding $\mathds{CP}^2 \hookrightarrow
\mathds{CP}^{N(N+3)/2}$ defines a manifold of $SU(3)$ coherent 
states. In terms of the usual spherical coordinates, we can write  
\begin{eqnarray}
(s,t,u)=(\sin\half\theta\cos\half\varphi,
\sin\half\theta\sin\half\varphi \re^{{\rm i}\xi}, \cos\half\theta
\re^{{\rm i}\eta}),
\end{eqnarray}
and substitution of this expression into the trinomial expansion 
for each $N$ then yields the parametric representation of the 
corresponding $\half(N^2+3N+2)$-level $SU(3)$ coherent state:
\begin{eqnarray}
\fl |\theta,\varphi,\xi,\eta\rangle = \sum_{k=0}^N \sum_{l=0}^k \!
\sqrt{ \! \textstyle{\frac{N!}{l!(N-k)!}}} \! \left(
\sin\half\theta\cos\half\varphi\right)^l \! \left(
\sin\half\theta\sin\half\varphi \re^{{\rm i}\xi}\right)^{k-l} \!
\left( \cos\half\theta \re^{{\rm i}\eta} \right)^{N-k} \!
|k,l\rangle.
\end{eqnarray}
The state space of $SU(3)$ coherent states may be regarded as
a classical phase space, and geometrically this is just a complex
projective plane $\mathds{CP}^2$ with curvature $6/N$ (see also 
Gnutzmann \& Ku\'s 1998 for a detailed analysis of the $SU(3)$ 
coherent states).

Similarly, for each $k$ we may consider the Veronese embedding
$\mathds{CP}^k \hookrightarrow \mathds{CP}^{k(k+3)/2}$ and its
generalisations. Then the validity of the following statement should 
be evident. 

\noindent \textbf{Proposition}. \textit{The totality of $SU(k+1)$ 
coherent states is characterised by the family of generalised 
Veronese varieties
\begin{eqnarray}
\mathds{CP}^k \hookrightarrow
\mathds{CP}^{\frac{1}{k!} (N+1)(N+2)\cdots(N+k)-1}, 
\qquad (N=2,3,\ldots). \label{eq:23}
\end{eqnarray}
Further, for each $k$ and $N$ the induced 
metric on the $SU(k+1)$ coherent state manifold is that of a 
$\mathds{CP}^k$ scaled by $N$, hence the curvature is
scaled by $1/N$.}
 
The standard spherical representation for 
$\mathds{CP}^k$ can thus be used to parameterise the coherent 
state manifolds, which may be useful in various applications 
such as passage to the classical limit (cf. Gnutzmann \& Ku\'s 1998; 
Graefe \textit{et al}. 2008; Trimborn \textit{et al}. 2008; Yaffe 1982) 
or, conversely, geometric quantisation (cf. Rawnsley 1977).

\vspace{0.4cm} 
\noindent 8. \textit{Coherent states on hyperbolic domains}. 
Counterparts of the $SU(k+1)$ coherent states, associated with 
the noncompact group $SU(1,k)$, also have various
applications in physics. We shall derive the metrics 
of the relevant state spaces by the methods outlined 
above. To this end, we consider a Hilbert space equipped 
with an indefinite inner product $\langle{\bar z},w\rangle_- = 
-{\bar z}_0w^0+{\bar z}_jw^j$ on ${\mathds C}^{k+1}$, 
with isometry group $SU(1,k)$. Restricting consideration to 
the subspace for which, say, $\langle{\bar z}, 
z\rangle_-<0$, and forming as before the space of rays through the 
origin, we obtain the complex hyperbolic space $\mathds{CH}^k$. 
Here, the analogue of the Hopf fibration $S^1\to S^{2k+1}\to 
\mathds{CP}^k$ is $S^1\to Q^{2k+1}\to\mathds{CH}^k$, where 
$Q^{2k+1}$ is the pseudo-sphere $\langle{\bar z}, z\rangle_-=-1$.
 
In the indefinite case, we have
\begin{eqnarray}
\frac{\langle {\bar z},z+\rd z\rangle_-  \langle {\bar z}+ \rd {\bar
z},z\rangle_-}{\langle {\bar z},z\rangle_-\langle {\bar z}+\rd{\bar
z},z+\rd z\rangle_-} = \cosh^2\half \rd s
\end{eqnarray}
in place of (\ref{eq:1}), and thus the metric on the
complex hyperbolic space $\mathds{CH}^k$ is
\begin{eqnarray}
\rd s^2 = -4 \frac{\langle {\bar z},z\rangle_- \langle \rd{\bar z},
\rd z\rangle_-  - \langle {\bar z},\rd z \rangle_-  \langle z,
\rd{\bar z} \rangle_-}{\langle {\bar z},z\rangle_-^2}. \label{eq:25}
\end{eqnarray}
In inhomogeneous coordinates, the hyperbolic metric (\ref{eq:25}) 
assumes the more familiar form (Kobayashi \& Nomizu 
1969)\footnote{Note the error in the sign of the
second term in the expression for the hyperbolic metric in 
Kobayashi \& Nomizu 1969.}:
\begin{eqnarray}
\rd s^2 =
4\frac{(1-{\bar\zeta}_j\zeta^j)(\rd{\bar\zeta}_j\rd\zeta^j)+
({\bar\zeta}_j\rd\zeta^j)(\zeta^j\rd{\bar\zeta}_j)}
{(1-{\bar\zeta}_j\zeta^j)^2}. \label{eq:26}
\end{eqnarray}
That (\ref{eq:26}) reduces to (\ref{eq:25}) can be verified by
substituting $\rd\zeta^j=(z^0\rd z^j-z^j\rd z^0)/(z^0)^2$ and
$\rd{\bar\zeta}^j=({\bar z}_0\rd {\bar z}_j-{\bar z}_j\rd {\bar
z}_0)/ ({\bar z}_0)^2$ into (\ref{eq:26}) and rearranging terms.
Alternatively, (\ref{eq:26}) may be deduced from the expression
$K=4\ln(1-{\bar\zeta}_j\zeta^j)$ for the K\"ahler potential. Up to a 
scale factor, (\ref{eq:26}) is the unique
Riemannian metric on $\mathds{CH}^k$ such that the action of
$U(1,k)$ is isometric.
 
Let us consider in more detail the geometry of the complex
hyperbolic line $\mathds{CH}^1$. The appropriate metric on this 
space is sometimes known as the Bergman metric, since it can be
obtained from of the Bergman kernel (Bergman 1970). We shall
briefly explain this idea in view of its relevance to the
theory of generalised coherent states. We have restricted the states
to the region where $\langle{\bar z}, z\rangle_-<0$, i.e. 
$-{\bar z}_0z^0+{\bar z}_1 z^1<0$. Passing to the
inhomogeneous coordinate, this inequality ${\bar\zeta}\zeta<1$ defines 
the unit (Poincar\'e) disk ${\mathcal D}$ in the complex plane. Consider
the Hilbert space $\mathcal{L}^2({\mathcal D})$ of complex-valued functions 
square-integrable with respect to the Lebesgue measure $\rd x\rd y$ 
on ${\mathcal D}$, where $\zeta=x+\ri y$, and the
complex orthonormal functions on ${\mathcal D}$ defined by
\begin{eqnarray}
\phi_n(\zeta) = \left(\frac{n}{\pi}\right)^{\frac{1}{2}} \zeta^{n-1}. 
\label{eq:27}
\end{eqnarray}
A direct calculation shows that $\{\phi_n\}$ forms an orthonormal 
basis for $\mathcal{L}^2({\mathcal D})$:
\begin{eqnarray}
\int\!\!\int_{{\mathcal D}} \rd x\, \rd y\, \overline{\phi_n(\zeta)}
\phi_m(\zeta) &=&  \frac{\sqrt{nm}}{\pi} \int_0^1 r \rd r
\int_0^{2\pi} \!\!\rd \theta\, r^{n+m-2} \re^{-{\rm i}\theta(n-m)}
\nonumber \\ &=& \frac{2\sqrt{nm}}{n+m}\, \delta_{nm} = \delta_{nm}.
\end{eqnarray}
The kernel $K_B$ corresponding to this orthonormal set is defined by 
\begin{eqnarray}
K_B(\zeta,\bar{\chi})=\sum_{n=1}^\infty\phi_n(\zeta)
\overline{\phi_n(\chi)} . \label{eq:28}
\end{eqnarray}
The function $K_B(\zeta,\bar{\chi})$, known as the
Bergman kernel, satisfies the identity
\begin{eqnarray}
g(\zeta)=\int\!\!\int_{{\mathcal D}}\rd x\,\rd
y\,K_B(\zeta,\bar{\chi}) g(\chi),
\end{eqnarray}
(where $\chi=x+\ri y$) for any smooth function
$g\in\mathcal{L}^2({\mathcal D})$. This kernel function generally 
diverges for a real orthonormal basis, but always converges 
uniformly in the complex case (Helgason 1978). In the present
case, substituting (\ref{eq:27}) into (\ref{eq:28}), we obtain
\begin{eqnarray}
K_B(\zeta,\bar{\chi})=\frac{1}{\pi(1-\zeta{\bar\chi})^2} .
\end{eqnarray}
 
The kernel function $K_B$ associated with a domain ${\mathcal
D}$ in the complex plane naturally determines a Riemannian
metric on ${\mathcal D}$, called the Bergman metric, via the
prescription
\begin{eqnarray}
\rd s^2= \frac{\partial^2}{\partial{\bar\zeta}\partial\zeta} \, \ln
K_B(\zeta,\bar{\zeta})\rd \bar{\zeta}\rd \zeta. \label{eq:metric}
\end{eqnarray}
Two salient properties of this metric are invariance
under conformal transformations and monotonicity in the
sense that if ${\mathcal D}'\subset {\mathcal D}$ then the 
associated line elements satisfy $\rd s'>\rd s$. In the present 
example this line element, up to a scale factor, is given by
\begin{eqnarray}
\rd s^2 = 4\, \frac{\rd{\bar\zeta}\rd\zeta}{(1-{\bar\zeta}\zeta)^2},
\label{eq:32}
\end{eqnarray}
which agrees with (\ref{eq:26}) for $k=1$.

The standard formulation of $SU(1,1)$ coherent states, due to
Solomon (1971) and Perelomov (1972), is  closely related to the 
geometric quantisation 
(cf. Odzijewicz 1992) of the Poincar\'e disk ${\mathcal D}$ defined by 
${\bar\zeta}\zeta<1$, and is based upon the infinite-dimensioal
Hilbert space ${\mathcal L}^2 ({\mathcal D})$ spanned by the
orthonormal functions (\ref{eq:27}). Specifically, Perelomov (1972)
defines, for any $|\xi|<1$, the generic coherent state as the Hilbert 
space vector
\begin{eqnarray}
|\xi\rangle = \sum_{n=1}^\infty \sqrt{n}\, 
\xi^{n-1} \phi_n(\zeta) 
\label{eq:x1x}
\end{eqnarray}
(here we consider the lowest-order state in
the representation). In this case, the Hilbert space ${\mathcal
L}^2({\mathcal D})$ is equipped with a positive-definite inner
product, which projectively defines the Fubini-Study metric. Moreover, 
 
\noindent \textbf{Theorem}. \textit{The metric of the
standard $SU(1,1)$ coherent state submanifold of the projective Fock
space induced by the ambient Fubini-Study metric is hyperbolic}.
 
\textit{Proof}. The Fubini-Study metric on the projective Hilbert space
assumes the form
\begin{eqnarray}
\rd s^2 = 4 \left[ \frac{\langle \rd\xi|\rd\xi\rangle}{\langle
\xi|\xi\rangle} - \frac{\langle \xi|\rd \xi\rangle \langle
\rd\xi| \xi\rangle} {\langle \xi|\xi \rangle^2} \right].
\label{eq:x2x}
\end{eqnarray}
For the coherent state (\ref{eq:x1x}), a simple calculation shows that
$\langle\xi|\xi\rangle = (1-{\bar\xi}\xi)^{-2}$,
$\langle\xi| \rd\xi\rangle = 2{\bar\xi}\rd
\xi(1-{\bar\xi}\xi)^{-3}$, and
$\langle\rd\xi|\rd\xi\rangle =
2(1+2{\bar\xi}\xi)(1-{\bar\xi}\xi)^{-4} \rd {\bar\xi}\rd
\xi$. Substituting these into (\ref{eq:x2x}), we find that the line
element is $\rd s^2=8(1-{\bar\xi}\xi)^{-2}
\rd{\bar\xi}\rd\xi$.  \hspace*{\fill} $\square$

This hyperbolic geometry of the $SU(1,1)$ coherent state space 
is well known (cf. Perelomov 1986), but to our knowledge its 
explicit derivation from the ambient Fubini-Study geometry, as 
illustrated here, has not previously appeared in the relevant 
literature. More generally, for ${\bar\xi}_j\xi^j<1$ we define the 
$SU(1,k)$-analogue of the $SU(1,1)$ coherent state (\ref{eq:x1x}) 
by
\begin{eqnarray}
|\xi\rangle = \left( 1, \sqrt{2}\xi^i, \sqrt{3}\xi^i\xi^j,
\sqrt{4}\xi^i \xi^j \xi^l, \cdots \right) .
 \label{eq:y1y}
\end{eqnarray}
Then, by a simple calculation, $\langle\xi|\xi\rangle =
(1-{\bar\xi}_j\xi^j)^{-2}$,  $\langle\xi|\rd\xi\rangle =
2(1-{\bar\xi}_j\xi^j)^{-3}{\bar\xi}_j\rd \xi^j$, and
$\langle\rd\xi|\rd\xi\rangle = 2(1-{\bar\xi}_j\xi^j)^{-3}
\rd {\bar\xi}_j\rd \xi^j + 6(1-{\bar\xi}_j\xi^j)^{-4}
({\bar\xi}_j\rd\xi^j)(\xi^j \rd{\bar\xi}_j)$. Substituting
these into (\ref{eq:x2x}) we obtain the line element 
\begin{eqnarray}
\rd s^2 =
8\frac{(1-{\bar\xi}_j\xi^j)(\rd{\bar\xi}_j\rd\xi^j)+
({\bar\xi}_j\rd\xi^j)(\xi^j\rd{\bar\xi}_j)}
{(1-{\bar\xi}_j\xi^j)^2}, \label{eq:26.5}
\end{eqnarray}
which is just twice the metric of the original $\mathds{CH}^k$.
 
\vspace{0.4cm} 
\noindent 9. \textit{Atomic coherent states for indefinite Hilbert 
spaces}. We shall now construct generalised coherent states 
associated with the group $SU(1,k)$ by defining Veronese-type 
maps for indefinite Hilbert spaces. This result provides 
finite-dimensional $SU(1,k)$-analogues of the $SU(k+1)$ 
coherent states defined on a $(k+1)$-dimensional Hilbert space 
with an indefinite Hermitian inner product of the Pontryagin type 
(Pontryagin 1944). The foregoing discussion of the Veronese 
embedding might convey the preliminary impression that the 
distinction between the state spaces $\mathds{CP}^k$ and 
$\mathds{CH}^k$ is merely formal and insignificant. However, 
there are essential differences between these two cases. In the 
present context, for instance, a Veronese construction for an 
embedding of the form, say, $\mathds{CH}^1 \hookrightarrow 
\mathds{CH}^{2}$ does not exist. To see this, let $(s,t)$ denote 
the homogeneous coordinates of $\mathds{CH}^1$. Then 
$-{\bar s}s+{\bar t}t=-1$; squaring this, we obtain $({\bar s}s)^2
-2 {\bar s}s{\bar t}t+({\bar t}t)^2=+1$. On the other, if $(s^2,
\sqrt{2}st, t^2)$ were the homogeneous coordinates of a point in 
$\mathds{CH}^{2}$, then we would have $-({\bar s}s)^2+2 
{\bar s}s{\bar t}t+({\bar t}t)^2=-1$, contradicting the previous 
equation. Alternatively, note that a point on $\mathds{CH}^{2}$ 
can be parameterised in the form 
\begin{eqnarray}
(z^0,z^1,z^2)=
\left(\cosh\half\tau,\sinh\half\tau \cos \theta 
\re^{{\rm i}\alpha},  \sinh\half\tau \sin\theta \re^{{\rm i}\beta}\right),
\end{eqnarray} 
whereas a point on $\mathds{CH}^{1}$ can be expressed in the 
form $(z^0,z^1)=(\cosh \frac{1}{2}\tau,\sinh \frac{1}{2} \tau 
\re^{{\rm i}\phi})$. Hence, $\mathds{CH}^1$ cannot be 
embedded into $\mathds{CH}^2$ via a Veronese-type
construction.
 
Nevertheless, one can define a Veronese-type map 
$\mathds{CH}^k \hookrightarrow {\mathfrak M}$ for a certain 
hyperbolic K\"ahler manifold ${\mathfrak M}$ having a signature 
structure distinct from that of any $\mathds{CH}^k$. For 
example, the embedding $(s,t) \hookrightarrow (s^3,\sqrt{3}
s^2t, \sqrt{3}st^2, t^3)$ of $\mathds{CH}^1$ defines an 
$SU(1,1)$ `coherent' state within the state space of an 
$SU(2,2)$ system, rather than an $SU(1,3)$ system. Since the 
state space of the $SU(2,2)$ system is not a $\mathds{CH}^{3}$, 
the interpretation of this embedding is somewhat different from 
that of the previously considered map relating to $SU(2)$ 
coherent states. Nevertheless, it is of interest, at least from a 
mathematical viewpoint, to formulate a concept of $SU(1,k)$ 
coherent states applicable to Hilbert spaces with indefinite 
inner products.  

We consider here only the case $k=1$. Using the standard
parameterisation $(s,t)=(\cosh \frac{1}{2}\tau,\sinh \frac{1}{2}
\tau \re^{{\rm i}\phi})$ for the homogeneous coordinates on 
$\mathds{CH}^1$, we obtain
\begin{eqnarray}
|\tau,\phi\rangle = \sum_{j=0}^{N} \sqrt{{\textstyle \left(
{N\atop j} \right)}} \left( \cosh\half \tau \right)^j \left(
\sinh\half \tau \re^{{\rm i}\phi} \right)^{N-j} |j\rangle ,
\label{eq:33}
\end{eqnarray}
which can be regarded as an embedding of $SU(1,1)$ `coherent' 
states within the state space of an $SU((N+1)/2,(N+1)/2)$ system 
if $N$ is odd, and an $SU(1+N/2,N/2)$ system if $N$ is even. For 
any $N$, the metric of this  submanifold can easily be calculated 
from the Bergman kernel, and the result is a hyperbolic metric of 
the form (\ref{eq:32}), scaled by the factor $N$. 

We find therefore that the construction of finite-dimensional 
$SU(1,1)$ coherent states is entirely feasible even though the 
standard algebraic definition precludes such an object because 
$SU( 1,1 )$ has no finite-dimensional unitary representations. 
This demonstrates the flexibility in our geometric construction of 
coherent state spaces. The coherent state (\ref{eq:33}) and its 
generalisations may prove useful in the various applications of 
the Pontryagin-Kre{\u\i}n spaces.

\vspace{0.4cm} 
\noindent 10. \textit{Discussion}. Our consideration
has been focussed upon the metric properties of the various coherent
state spaces; the algebraic geometry of these spaces is rather
intricate and will be discussed
elsewhere. Here, we merely mention that the $SU(k+1)$ coherent states for
$k=1,2,\ldots$ possess a natural ``hierarchical" structure arising from the 
Veronese subvarieties associated with a series of embeddings of the form
\begin{eqnarray}
\mathds{CP}^1 \hookrightarrow \mathds{CP}^{2} \hookrightarrow
\mathds{CP}^{5} \hookrightarrow \mathds{CP}^{20} \hookrightarrow
\mathds{CP}^{230} \hookrightarrow \cdots
\end{eqnarray}
for $SU(2)$ and $SU(3)$, and its generalisations (e.g.,
$\mathds{CP}^3 \hookrightarrow \mathds{CP}^{9} \hookrightarrow
\mathds{CP}^{54} \hookrightarrow \mathds{CP}^{1539} \hookrightarrow
\cdots$ for $SU(4)$; $\mathds{CP}^4 \hookrightarrow \mathds{CP}^{14}
\hookrightarrow \mathds{CP}^{119} \hookrightarrow \mathds{CP}^{7497}
\hookrightarrow \cdots$ for $SU(5)$, and so on). Thus, for example,
within the space of $SU(3)$ coherent states corresponding to each value
of $N$ there is a subanifold of $SU(2)$ coherent states, and so on.
By means of generalised Veronese embeddings, the
nesting of these $SU(k+1)$ coherent states can be succinctly 
described by elementary combinatorics. In particular, the
natural metric structures of all these subspaces are of the
Fubini-Study type.
 
\vskip 10pt The authors thank D.~Blasius, E.~J.~Brody and 
L.~P.~Hughston for useful comments and stimulating discussions.

\end{document}